\documentclass[preprintnumbers,amsmath,amssymb,floatfix,11
pt,prd,superscriptaddress,nofootinbib]{revtex4}
\usepackage{graphicx}
\usepackage{epsfig}
\usepackage{bm}
\usepackage{amsfonts}

\input epsf

\begin{document}

\title{\textbf{Holographic reconstruction of k-essence model with Tsallis and the most generalized Nojiri-Odintsov version of holographic dark energy}
}%

\author{Ulbossyn Ualikhanova}
\email{ulbossyn.ualikhanova@gmail.com}
\author{Aziza Altaibayeva}
\email{aziza.ltaibayeva@gmail.com}
\affiliation{ Department of General and Theoretical Physics, L. N. Gumilyov Eurasian National University, Satpayev Str. 2, 010008 Astana, Kazakhstan.}

\author{Surajit Chattopadhyay\footnote{Communicating author}}
\email{schattopadhyay1@kol.amity.edu; surajitchatto@outlook.com}
\affiliation{ Department of Mathematics, Amity University, Kolkata, Major Arterial Road, Action Area II, Rajarhat, Newtown, Kolkata 700135,
India.}

\date{\today}% It is always \today, today,

\begin{abstract}
\begin{center}
    \bf{Abstract}
\end{center}
The holographic principle, which has its roots in string theory and black hole thermodynamics, connects the maximum distance of a quantum field theory to its infrared cutoff, which is correlated with the vacuum energy. The present study explores a reconstruction scheme for the k-essence form of dark energy with the most generalized version of holographic dark energy introduced in S. Nojiri, and S. D. Odintsov (2006) (Gen. Relativ. Gravit., 38 p: 1285-1304 ) and (S. Nojiri and S. D. Odintsov, 2017, European Physical Journal C, 77, pp.1-8 ). Here, in the initial phase of the study, we begin with a reconstruction scheme of the k-essence model with Tsallis holographic dark energy and finally with a highly generalized version of holographic dark energy with Nojiri-Odintsov generalization. Finally, we have studied the cosmological consequences of the k-essence dark energy with the generalized versions of holographic fluid.
\end{abstract}

%Use showkeys class option if keyword
\maketitle

%\tableofcontents

\section{Introduction}\label{sec1}

Many cosmological experiments, such as observations of large-scale structures, searches for type-Ia supernovae, and cosmic microwave background anisotropy measurements \cite{obs1, obs2}, all show that the universe's expansion is accelerating \cite{nojiri1, nojiri2, nojiri3, acceleration1, acceleration2, acceleration3}. If the universe contains ideal fluids, the equation of state (EoS) parameter $w=p/\rho$ (equal to the ratio of its pressure $p$ to its energy density $\rho$) must be less $-\frac{1}{3}$, suggesting negative pressure. Several ideas have been developed to explain the negative pressure or repulsive gravitational behavior. One theory suggests the existence of ``dark energy" (DE), smooth energy with negative pressure \cite{DE1, DE2, DE3, DE4}. DE can be attributed to the dynamics of a scalar field, such as the quintessence \cite{quintessence1}. Late-time attractor solutions, replicating the perfect fluid in various parameters, successfully explain the present-day cosmic acceleration. Other scalar field models that serve as DE models include phantom, quintom, tachyon, k-essence, dilaton, hessence, Dirac–Born–Infeld (DBI)-essence, pilgrim, and others (for reviews, see \cite{acceleration1, acceleration2, acceleration3, rami1, rami2, rami3}). Correspondences between dark energy theories with scalar fields and changed gravities remain problematic in cosmology. Several authors have explored the relationship between different DE models and their cosmological implications. Previously, some authors investigated Generalized Ghost Pilgrim Dark Energy (GGPDE) models. Some recent works are available, where the relationship between GGPDE and scalar field dark energy has been investigated \cite{pde1, pde2, pde3}. Pasqua and Chattopadhyay \cite{new1} proposed an effective scalar field theory characterized by a Lagrangian with a noncanonical kinetic term, which leads to rapid expansion in the present Universe and is known as k-essence in the context of El-Nabulsi's fractional action cosmology \cite{new2, new3}.

A comprehensive quantum theory of gravity can help accurately measure vacuum energy \cite{HDE1}. Although we lack such a comprehensive theory, one can examine the nature of DE using quantum gravity concepts. The holographic concept, in particular, is an important feature that may aid in resolving cosmological and DE concerns. The reference \cite{HDE1} proposed the holographic dark energy (HDE) and new agegraphic dark energy (NADE) models as equal descriptions of DE components derived from the modified theory. Some analytical solutions were achieved by \cite{HDE1} applying the beginning conditions to the corresponding functions. They \cite{HDE1} determined the explicit $f(R, T)$ functions for HDE and NADE. The reference \cite{HDE2} proposed a holographic description of four-dimensional single-scalar inflationary universes and demonstrated how three-dimensional quantum field theory correlation functions encapsulate cosmic observables like the primordial power spectrum. They \cite{HDE2} also show that the holographic description accurately reproduces traditional inflationary predictions in the domain where perturbative quantization of fluctuations is justified. Using the HDE model with the Hubble horizon as the infrared cut-off, reference \cite{HDE3} reconstructed the interaction rate between dark energy and dark matter for three different parameterizations of the deceleration parameter. The model parameters were obtained \cite{HDE3} by maximum likelihood analysis using observational Hubble parameter data (OHD), type Ia supernova data (SNe), baryon acoustic oscillation data (BAO), and the distance before the cosmic microwave background.

In a noteworthy work, Nojiri et al. \cite{Odi1} employed the holographic approach to represent our universe's early-time and late-time acceleration eras in a unified fashion. They discovered a correlation with several higher curvature cosmological models with or without matter fields. The holographic cutoffs were determined \cite{Odi1} in terms of the particle or future horizon and its derivatives. As a result, they \cite{Odi1} proposed the holographic energy density to combine multiple cosmological epochs of the universe from a holographic point of view. In another critical work, \cite{Odi2} investigated numerous entropic dark energy models—such as the Tsallis entropic DE, the Rényi entropic DE, and the Sharma-Mittal entropic DE—and concluded that they are undoubtedly equivalent to the generalized HDE. Odintsov et al. \cite{Odi3} studied the universe's early evolution, from inflation to reheating through holographic fluid, in entropic cosmology, where the entropic energy density acts as the inflaton. They \cite{Odi3} observed that the entropic energy successfully drives an early inflationary phase with a graceful exit.

Moreover, the theoretical expectations of the observable indices are consistent with the recent Plank data in \cite{Odi3}. Guberina et al. \cite{guberina} used a holographic DE model with variable cosmological-constant energy density and Newton constant to investigate the problem of determining an effective field-theory IR cutoff. Assuming ordinary matter scales canonically, \cite{guberina} demonstrated that the continuity equation unambiguously determines the IR cutoff when a law of variation for the energy density or Newton constant is known. In the generalized HDE formalism proposed by Nojiri et al. \cite{Odi4}, the holographic cut-off can be expressed as $L_{IR}=L_{IR}(L_p,\dot{L}_P, \ddot{L}_P,...,L_f,\dot{L}_f, \ddot{L}_f,...,a) $, where $L_p$ and $L_f$ represent the particle and future horizons, respectively, where $a$ represents the universe's scale factor. Chattopadhyay et al. \cite {binu1} studied a DE model in which energy density is a function of the Hubble parameter $H$ squared and of the first and second-time derivatives with respect to the cosmic time $t$ of $H$. With this version of HDE model, a specific case of the Nojiri-Odintsov holographic DE \cite{Odi5}, they \cite {binu1} established a correspondence between the generalized version of the holographic dark energy model and the tachyon, k-essence, dilaton, and quintessence scalar field models. Chakraborty et al. \cite{binu2} investigated the cosmology and baryogenesis of modified $f(G)$ gravity, assuming the background evolution as generalized holographic dark energy. In another work, \cite{binu3} proposed a way to reconstruct a generalized form of holographic dark energy and analyze its cosmology based on a logotropic equation of state, assuming that the dark energy density is caused by holographic Ricci dark energy. In the framework of entropic cosmology, in which the entropic energy density serves as the inflaton, \cite{Odi6} studied the evolution of the universe in its early stages, specifically from inflation to reheating.

In the present study, we aspire to see the cosmological consequences of reconstructing k-essence model of dark energy with some generalized versions of holographic fluid. For that purpose, we have carried out the reconstruction schemes for k-essence with Tsallis holographic dark fluid with different kinds of IR cut-offs. In the next phase, we considered the most generalized Nojiri-Odintsov holographic dark fluid and studied the cosmological consequences accordingly. The rest of the paper is organized as follows: Section II presented an overview of Tsallis holographic dark energy (THDE) and demonstrated the reconstruction of k-essence with different IR cut-offs. Section III discussed the reconstruction of the k-essence scalar field with the most generalized Nojiri-Odintsov holographic dark energy. We have concluded in Section IV.

\section{Tsallis holographic dark energy: An overview}

Holography's cosmological applicability relies on the notion that the universe's entropy is proportional to its area, similar to that of a black hole. In 1902, Gibbs noted that the Boltzmann-Gibbs theory could not be applied to systems with divergent partition functions, including gravitational systems \cite{revision8}. Tsallis demonstrated that in such instances, the Boltzmann-Gibbs additive entropy (based on weak probabilistic correlations and ergodicity) must be applied to non-additive entropy. The matter was elaborated in detail by Saridakis et al. \cite{saridakis1}. Tsallis' notion of entropy is critical for investigating gravitational and cosmic systems within generalized statistical mechanics. Tsallis and Cirto \cite{tsallis1} demonstrated that applying the Tsallis statistics to the system yields more than just the Bekenstein entropy. Quantum gravity confirms that the Tsallis entropy content of a system follows a power-law relationship with its area \cite{tsallis2}. In this context, let us mention that the standard holographic density is given by $\rho_D=3c^2 M_p^2 L^{-2}$, and it depends on the entropy–area relationship. Tsallis and Cirto \cite{tsallis1} showed that the horizon entropy of a black hole may be modified as follows:
\begin{equation}\label{1}
S_{\delta}=\gamma A^{\delta}
\end{equation}
Where $A=4\pi L^2$ represents the area of the horizon. $\gamma$ is an unknown constant and $\delta$ denotes the non-additivity parameter.  Bekenstein entropy is recovered at the appropriate limit of $\delta=1$ and $\gamma=1/4G$. Quantum gravity confirms this relationship, with implications for cosmology and holography \cite{tsallis2,  tsallis3, tsallis4, tsallis5, tsallis6}. According to the holographic principle, a physical system's degrees of freedom should scale with its enclosing area rather than its volume \cite{hp1, hp2, hp3, hp4}. The system entropy $(S)$ is related to the IR $(L)$ and UV $(\Lambda)$ cut-off is given by \cite{tsallis2}
\begin{equation}\label{2}
L^3\Lambda^3 \leq (S_{\delta})^{3/4}
\end{equation}
Using Eq. (\ref{1}) in (\ref{2}) and using some simplification one can get \cite{tsallis2}
\begin{equation}\label{3}
\Lambda^4 \leq (\gamma (4 \pi)^\delta) L^{2\delta-4}
\end{equation}
Here, $\Lambda^4$ represents the vacuum energy density and DE $\rho_{DE}$ energy density in the HDE hypothesis \cite{tsallis2}. References \cite{tsallis2,  tsallis3, tsallis4, tsallis5, tsallis6} proposed the Tsallis holographic dark energy density (THDE) based on the abovementioned inequality. The energy density of THDE is \cite{tsallis2}
\begin{equation}\label{4}
\rho_{DE}=B L^{2\delta-4}
\end{equation}
In this case, $B$ is an unknown parameter.

\subsubsection{The case of $L=H^{-1}$}
Following \cite{tsallis2}, let us consider a flat FRW universe for which the Hubble horizon is a proper candidate for the IR cutoff, and there is no interaction between the DE candidate and other components of the universe. Thus, in the present consideration $L=H^{-1}$. Using this in Eq. (\ref{4}) we have
\begin{equation}\label{5}
\rho_{DE}=B H^{-2\delta+4}
\end{equation}
In this non-interacting scenario, denoting the EoS parameter by $w_{DE}$, we can write the conservation equation as
\begin{equation}\label{6}
\dot{\rho}_{DE}+3H \rho_{DE} (1+w_{DE})=0
\end{equation}
Using Eq. (\ref{5}) in (\ref{6}) we obtain
\begin{equation}\label{7}
w_{DE}=-1+\frac{2(-2+\delta)\dot{H}}{3H^2}
\end{equation}

\subsubsection{The case of $L=(\alpha H^2+\beta \dot{H})^{-1/2}$}
In case we consider $L=(\alpha H^2+\beta \dot{H})^{-1/2}$ \cite{revision9} the dark energy density (\ref{5}) takes the form
\begin{equation}\label{8}
\rho_{DE}=B(\alpha H^2+\beta \dot{H})^{-\delta+2}
\end{equation}
The Hubble parameter $(H =\frac{\dot{a}}{a})$ and constants $(\alpha)~\text{and}~(\beta)$ must adhere to existing observational data constraints. The origin of holographic dark energy remains unknown, although the inclusion of the time derivative of the Hubble parameter is expected as it exists in the curvature scalar.
Using (\ref{8}) in (\ref{6}), we obtain the reconstructed EoS parameter as follows:

\begin{equation}\label{thdew}
    w_{DE}=\frac{-3 \alpha  H^3+\left(-3 \beta +4 \alpha  (-2+\delta )^2\right) H \dot{H}+2 \beta  (-2+\delta )^2 \ddot{H}}{3 \left(\alpha  H^3+\beta
 H \dot{H}\right)}
\end{equation}

\begin{table}[h!]
\caption{Current values $(z=0)$ of $w_{DE}$ based on Eq. (\ref{thdew}). Values are compared with current EoS values $-1.06_{-0.13}^{+0.11}$ from observational data sets from SNLS3, BAO and
Planck + WMAP9 + WiggleZ available in \cite{revision2}.}
\begin{center}
\begin{tabular}{||c c c c||}
\hline
 $(A,b,B,C_1)$ & $n=0.01$ & $n=0.001$ & $n=0.0001$ \\ [0.5ex]
 \hline\hline
 $(0.3,0.6,0.3,2)$ & $-0.991484$ & $-1.01472$ & -1.01714 \\
 \hline
 $(0.4,0.6,0.4,1.9)$ & $-1.02019$ & $-1.03718$ &$-1.03892$ \\
 \hline
 $(0.4,0.5,0.4,3)$ & $-1.00546$ & $-1.02757$ & $-1.02987$ \\
 \hline
 $(0.4,0.4,0.3,1.3)$ & $-1.00618$ & $-1.01398$ & $-1.01477$ \\
 \hline
$(0.1,0.4,0.5,2.8)$ &$-0.933559$ & $-0.993365$ & $-1.00001$ \\ [1ex]
 \hline
\end{tabular}
\label{table:1}
\end{center}
\end{table}

Reference \cite{revision2} employed Chevallier-Polarski-Linder parametrization to show that current/future measurements support the CDM scenario over the WDM for dark energy EoS. They constrained the current value of the EoS parameter by $-1.06_{-0.13}^{+0.11}$ from observational data sets from SNLS3, BAO and $Planck + WMAP9 + WiggleZ$. In a more recent study \cite{revision3}, the present value of DE EoS is restricted by observations to be close to $-(-1.019_{-0.080}^{+0.075})$ according to the $95\%$ limits imposed by Planck data combined with other astrophysical measurements \cite{revision4}.

\subsection{Reconstruction of the k-essence model}
This section will discuss the scalar field and potential associated with the k-essence model and rebuild them using the correspondence with the THDE in the FRW backdrop. The k-essence model explains the universe's late-time acceleration \cite{kessence1,kessence2,kessence3}. K-essence situations exhibit attractor-like dynamics, eliminating the need to tune the scalar field's beginning conditions individually. The approach adopted here is inspired by the work of \cite{kessence1}. This kind of model is characterized by non-standard kinetic energy terms and is described by a general scalar field action, a function of $\phi$ and $X=-\frac{1}{2} \partial_{\mu}\phi \partial^{\mu} \phi$. The action is given by \cite{kessence4, kessence5}

\begin{equation}
    S=\int d^4x \sqrt{-g}p(\phi,X)
\end{equation}

where $p(\phi; X)$ corresponds to a pressure density and usually is restricted to the Lagrangian density of the form $p(\phi, X)=f(\phi) g(X)$. The energy-momentum tensor for this Lagrangian density is used to calculate the energy density of the field (see below) \cite{kessence4, kessence5};

\begin{equation}\label{rho}
    \rho(\phi, X)=f(\phi) (-X+3X^2)
\end{equation}

Lagrangian density can be transformed into the pressure as follows:

\begin{equation}\label{p}
    p(\phi, X)=f(\phi) (-X+X^2)
\end{equation}

Using Eqs. (\ref{rho}) and (\ref{p}), we obtain the equation of state parameter

\begin{equation}\label{w}
   w_X=\frac{X-1}{3X-1}
\end{equation}

As we carry out a reconstruction scheme, we consider a correspondence $w_{DE}=w_X$, and $X$ gets the holographically reconstructed form

\begin{equation}\label{w}
   X=\frac{3H^2+(2-\delta)\dot{H}}{6H^2+3(2-\delta)\dot{H}}
\end{equation}

which, on further simplification, gives

\begin{equation}\label{14}
  X=\frac{1}{3}+\frac{H^2}{6 H^2-3 (-2+\delta ) \dot{H}}
\end{equation}

Clearly, from Eq. (\ref{w}), it is understandable that for $1/2<X<2/3$, we can have $-1<w_X<-(1/3)$, which is necessary for the accelerated expansion of the universe. From this, we can derive the following:

\begin{equation*}
    \frac{\dot{H}}{H^2}<(-2+\delta)^{-1}
\end{equation*}

Now, let us consider Eq. (\ref{thdew}), where the EoS THDE is reconstructed by taking $L=(\alpha H^2+\beta \dot{H})^{-1/2}$. As we are considering the holographic k-essence model, let us consider

\begin{equation}\label{wX}
X=\frac{1-w_{DE}}{1-3w_{DE}}
\end{equation}

where, $w_{DE}$ comes from  Eq. (\ref{thdew}) and $X$ becomes

\begin{equation}\label{17}
    X=\frac{3 \alpha H^3+\left(3 \beta -2 \alpha  (-2+\delta )^2\right)H \dot{H}-\beta  (-2+\delta )^2 \ddot{H}}{6 \alpha H^3+6 \left(\beta
-\alpha  (-2+\delta )^2\right)H \dot{H}-3 \beta  (-2+\delta )^2 \ddot{H}}
\end{equation}

Eqs. (\ref{14}) and (\ref{17}) are two holographically reconstructed forms of $X$ based on two different forms of IR cutoffs.

Let us now consider the scale factor to be in emergent form \cite{emergent}. An emergent universe model is intriguing to create, and it has the potential to address a few fundamental flaws with the classic Big Bang model. According to the emergent scenario, the Einstein static universe is an asymptote for time traveling \cite{emergent}. Starting from this phase, the standard expansion phase begins, avoiding the big bang singularity. The form of the scale factor is \cite{emergent}
\begin{equation}\label{scale}
    a(t)=A (e^{nt}+B)^b
\end{equation}
where, $A,~n,~B,~b$ are constants.

To create a model of the emergent cosmos, we make the following assumptions \cite{revision7}.
\begin{enumerate}
    \item The universe is isotropic and homogeneous on a wide scale.
    \item WMAP measurements suggest that it is spatially flat, with a total density parameter of $t = 1.02 \pm 0.02$.
    \item It is ever-present. There is no singularity.
    \item The universe is always vast enough that a classical description of spacetime is sufficient.
    \item Quantum field theory must be used to describe the source of gravity.
    \item The universe may contain exotic matter that violates energy requirements.
\end{enumerate}
We choose $A>0,~B>0$ so that the initial singularity is avoided and the scale factor is positive. Hence, $H=\frac{bne^{nt}}{B+e^{nt}}$. Accordingly Eqs. (\ref{14}) and (\ref{17}) takes the following forms:

\begin{equation}\label{case1}
    X_{case 1}=\frac{-3 b e^{n t}+B (-2+\delta )}{-6 b e^{n t}+3 B (-2+\delta )}
\end{equation}
and
\begin{equation}\label{case2}
    X_{case 2}=\frac{3 b^2 e^{2 n t} \alpha -b B e^{n t} \left(-3 \beta +2 \alpha  (-2+\delta )^2\right)+B \left(-B+e^{n t}\right) \beta  (-2+\delta
)^2}{6 b^2 e^{2 n t} \alpha -6 b B e^{n t} \left(-\beta +\alpha  (-2+\delta )^2\right)+3 B \left(-B+e^{n t}\right) \beta  (-2+\delta )^2}
\end{equation}

\begin{figure}
    \centering
    \includegraphics[width=0.6\linewidth]{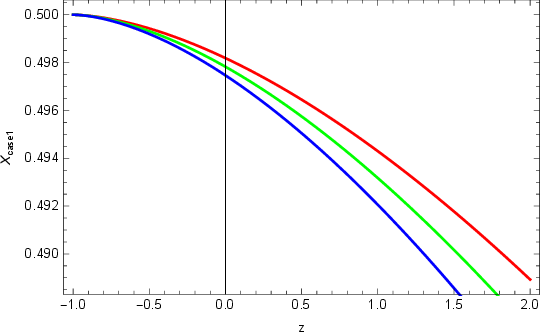}
    \caption{Evolution of $X$ with redshift $z$ as derived in Eq. (\ref{case1}) based on (\ref{14}). The red, green, and blue lines correspond to $B=0.5,~0.6~\text{and}~0.6$, respectively.}
    \label{fig:1}
\end{figure}

\begin{figure}
    \centering
    \includegraphics[width=0.6\linewidth]{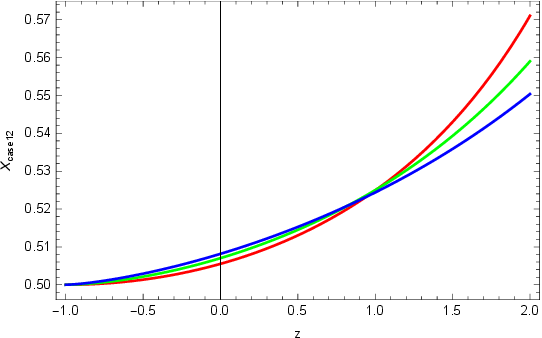}
    \caption{Evolution of $X$ with redshift $z$ as derived in Eq. (\ref{case2}) based on (\ref{14}). The red, green, and blue lines correspond to $B=0.5,~0.6~\text{and}~0.6$, respectively.}
    \label{fig:2}
\end{figure}

\begin{figure}
    \centering
    \includegraphics[width=0.6\linewidth]{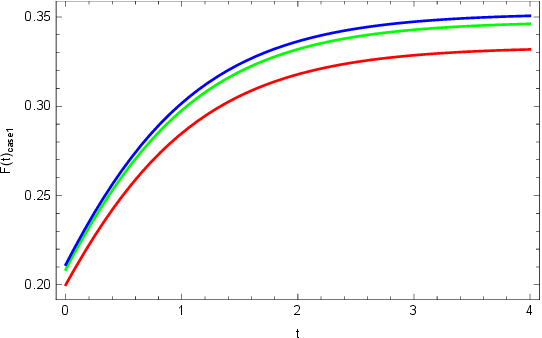}
    \caption{Evolution of $\mathcal{F} (t)_{case 1}$ with $t$ as derived in Eq. (\ref{f1}). The red, green, and blue lines correspond to $n=1.2,~1.213~\text{and}~1.217$, respectively.}
    \label{fig:3}
\end{figure}

\begin{figure}
    \centering
    \includegraphics[width=0.6\linewidth]{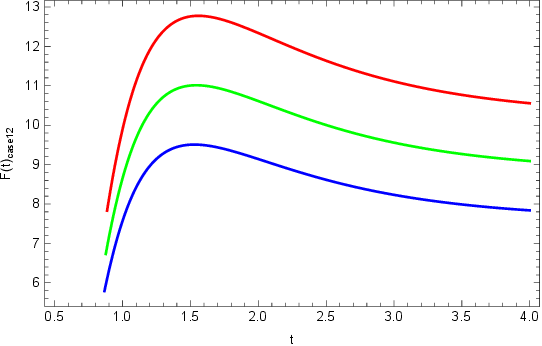}
    \caption{Evolution of $\mathcal{F} (t)_{case 2}$ with $t$ as derived in Eq. (\ref{case2}). The red, green, and blue lines correspond to $n=1.02,~1.03~\text{and}~1.04$, respectively.}
    \label{fig:4}
\end{figure}

In Fig. \ref{fig:1}, we have plotted $X$ as obtained in Eq. (\ref{case1}), which is the holographically reconstructed $X$, where we have chosen $L=H^{-1}$. In this figure, it is showing an increasing pattern with redshift $z$. This means that with the evolution of the universe, $X$ is increasing in this case.  On the contrary, in Fig. \ref{fig:2}, where we have plotted $X$ against $z$ based on Eq. (\ref{case2}), the $X$ is showing a decaying pattern with the evolution of the universe. In creating the figures, we have taken $A=0.3,~\delta=0.05,~b=0.6,~\text{and}~n=0.4$. Considering k-essence holographically, we can consider $\rho_{DE}=\rho(\phi, X)=f(\phi) (-X+3X^2)$. Thus, $f(\phi)=\frac{\rho_{DE}}{(-X+3X^2)}$. Subsequently, we reconstruct $f(\phi)$ as function of $t$ and denote it by $\mathcal{F} (t)$ and we get

\begin{equation}\label{f1}
\mathcal{F} (t)_{case 1}=\frac{3 b^3 B e^{3 n t} \left(\frac{1+B e^{-n t}}{b n}\right)^{2 \delta } n^4 \left(-2 b e^{n t}+B (-2+\delta )\right)^2}{\left(B+e^{n
t}\right)^4 \left(3 b e^{n t}-B (-2+\delta )\right)}
\end{equation}
and
\begin{equation}\label{f2}
\begin{array}{c}
 \mathcal{F} (t)_{case 2}= \left(\frac{3 B e^{-9 n t} \left(B+e^{n t}\right)^{16} \left(\frac{b e^{n t} n^2 \left(b e^{n t} \alpha +B \beta \right)}{\left(B+e^{n t}\right)^2}\right)^{4
\delta } \left(\left(\frac{b e^{n t} n^2 \left(b e^{n t} \alpha +B \beta \right)}{\left(B+e^{n t}\right)^2}\right)^{2-\delta }\right)^{2 \delta }
}{b^9
n^{16} \left(b e^{n t} \alpha +B \beta \right)^9 \left(3 b^2 e^{2 n t} \alpha -b B e^{n t} \left(-3 \beta +2 \alpha  (-2+\delta )^2\right)+B \left(-B+e^{n
t}\right) \beta  (-2+\delta )^2\right)}\right) \times \\
\left(2 b^2 e^{2 n t} \alpha -2 b B e^{n t} \left(-\beta +\alpha  (-2+\delta )^2\right)+B \left(-B+e^{n t}\right) \beta  (-2+\delta )^2\right)^2
\end{array}
\end{equation}

In Figs. \ref{fig:3} and \ref{fig:4} we have plotted $\mathcal{F} (t)_{case 1}$ and $\mathcal{F} (t)_{case 2}$ with $t$. In case of $\mathcal{F} (t)_{case 1}$ we observe that $\mathcal{F} (t)_{case 1} \rightarrow 0$ as $t \rightarrow 0$. This indicates satisfaction of a sufficient condition for a realistic reconstruction model. Now, let us look at Fig. \ref{fig:4}. We observe that the reconstructed $\mathcal{F} (t)$ initially has an increasing pattern with $t$, but after a certain stage, it shows a decaying pattern with $t$. However, $\mathcal{F} (t)_{case 2} \rightarrow 0$ as $t \rightarrow 0$ is satisfied in this case also. Thus, one of the conditions for a realistic model is satisfied here.

\section{k-essence model with Nojiri-Odintsov holographic dark energy}

Reference \cite{Odi6} has demonstrated that the coincidence problem can be overcome in generalized holographic dark energy. A similar scenario was presented in \cite{Odi7}, where the interaction between matter and holographic energy allows for a constant ratio of matter-energy density to holographic energy. The ratio can remain constant if the effective EoS varies over time. The ratio of matter-energy density to dark energy density generally varies \cite{Odi8}. The holographic cut-off is commonly associated with the particle horizon or future horizon. In \cite{Odi4, Odi6}, it was postulated that the holographic cut-off (LIR) is influenced by particle and future event horizons, including their derivatives \cite{Odi4}
\begin{equation}\label{23}
    L_{IR} = L_{IR}\left(L_p, \dot{L}_p, \ddot{L}_p,··· , L_f, \dot{L}_f,··· ,a\right)
\end{equation}
Here, $L_p$ and $L_f$ are the particle horizon and future
horizon, respectively, and $a$ is the scale factor of the universe. Using this formalism, \cite{Odi4} demonstrated that the Barrow entropic dark energy DE model is identical to the generalized HDE, where the holographic cut-off is defined by the particle horizon and derivative and the future horizon and derivative. Such a generalized version of HDE, introduced by Nojiri and Odintsov in \cite{Odi6}, leads to
interesting phenomenology, both from inflation and dark energy perspective \cite{Odi1}. In this study, such a generalized version of HDE would be referred to as Nojiri-Odintsov holographic dark energy (NO-HDE). We will consider, inspired by \cite{martiros1}, this highly generalized model of holographic dark energy with the Nojiri-Odintsov cut-off is defined as follows \cite{martiros1}:
\begin{equation}
    \rho_{NO-HDE}=\frac{3c^2}{L^2}
\end{equation}
where,
\begin{equation}
    \frac{c}{L}=\frac{1}{L_f}\left[\alpha_0+\alpha_1 L_f+\alpha_2 L_f^2\right]
\end{equation}
where $L_f$ is the future horizon and is defined as
\begin{equation}
    L_f=a \int_{t}^{\infty}\frac{dt}{a}
\end{equation}
For the scale factor given in Eq. (\ref{scale}), we have the Hubble parameter in the following form:
\begin{equation}\label{Hnew}
    H=\frac{b e^{n t} n}{B+e^{n t}}
\end{equation}

Using $\dot{L_f}=HL_f-1$, we obtain for the scale factor (\ref{scale}) and subsequent Hubble parameter (\ref{Hnew}) the following solution for $L_f$
\begin{equation}\label{lf}
    L_f=\frac{1}{b n} \left(1+B e^{-n t}\right) \text{2F1}\left[1,1,1+b,-B e^{-n t}\right]+\left(B+e^{n t}\right)^b C_1
\end{equation}
Eq. (\ref{lf}) is regarded as the Nojiri-Odintsov cutoff. Using this cutoff, we have the Nojiri-Odintsov holographic dark energy (NO-HDE) density as
\begin{equation}\label{nohde}
 \rho_{NO~HDE}=\frac{3 b^2 c^2 n^2}{\left(b \left(B+e^{n t}\right)^b n C_1+\left(1+B e^{-n t}\right) \text{2F1}\left[1,1,1+b,-B e^{-n
t}\right]\right)^2}
\end{equation}
From Eq.(\ref{nohde}), we have the time derivative
\begin{equation}\label{nohdedot}
  {\dot \rho_{NO~HDE}}  =-\left[\frac{6 b^3 c^2 n^3 \left(-1+b e^{n t} \left(B+e^{n t}\right)^{-1+b} n C_1+2F1\left[1,1,1+b,-B e^{-n t}\right]\right)}{\left(b
\left(B+e^{n t}\right)^b n C_1+\left(1+B e^{-n t}\right) 2F1\left[1,1,1+b,-B e^{-n t}\right]\right)^3}\right]
\end{equation}
Using Rqs. (\ref{Hnew}), (\ref{nohde}) and (\ref{nohdedot}) in the conservation equation of the non-interacting scenario, the EoS parameter for the NO-HDE takes the form

\begin{figure}
    \centering
    \includegraphics[width=0.5\linewidth]{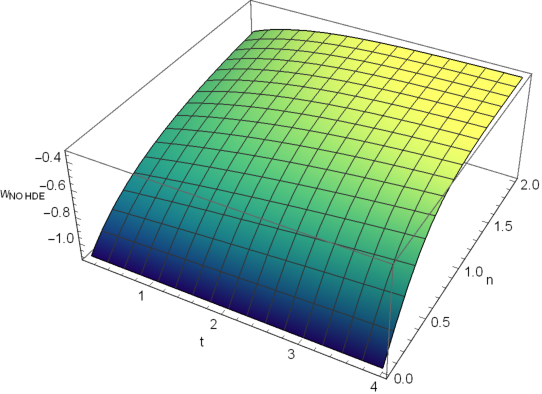}
    \caption{Evolution of $w_{NO HDE}$ for a range of values of $n$ based on Eq. (\ref{EoSNOHDE}).  }
    \label{fig:5}
\end{figure}
\begin{equation}\label{EoSNOHDE}
    w_{NO HDE}=-1+\frac{2}{3}\left(1-\frac{2}{3 \left(b e^{n t} \left(B+e^{n t}\right)^{-1+b} n C_1+2F1\left[1,1,1+b,-B e^{-n t}\right]\right)}\right)
\end{equation}
The EoS parameter for the NO-HDE obtained in (\ref{EoSNOHDE}) is plotted in Fig. \ref{fig:5} for a range of values of $n$. The EoS parameter is observed to be close to $-1$ for the values of $n$ close to $0$. However, for larger values of $n$, the EoS parameter behaves like a quintessence.

As we are considering the k-essence model of dark energy in the Nojiri-Odintsov holographic framework, we have from
\begin{equation}
    X=\frac{w_{NO HDE}-1}{3 w_{NO HDE}-1}
\end{equation}
that
\begin{equation}\label{Xnohde}
    X_{NO HDE}=\frac{1}{3} \left(2-\frac{1}{1+2 \text{F1}\left[1,1,1+b,-B e^{-n t}\right]+b e^{n t} \left(B+e^{n t}\right)^{-1+b} n C_1}\right)
\end{equation}

\begin{figure}
    \centering
    \includegraphics[width=0.5\linewidth]{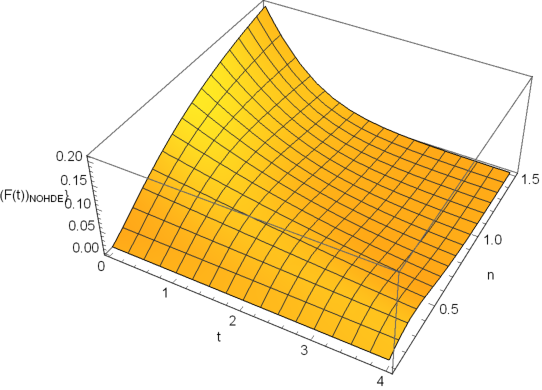}
    \caption{Evolution of $\mathcal{F}\left(t\right)_{NOHDE}$ based on Eq. (\ref{Fnohde}) for a range of values of $n$.}
    \label{fig:6}
\end{figure}

In this cosmological framework, we reconstruct $f(\phi(t))=\mathcal{F}(t)_{NOHDE}$ as
\begin{equation}\label{Fnohde}
\begin{array}{cc}
\mathcal{F}\left(t\right)_{NOHDE}=   \left( \frac{9 b^2 c^2 n^2}{\left(b \left(B+e^{n t}\right)^b n C_1+2 \left(1+B e^{-n t}\right) \text{F1}\left[1,1,1+b,-B e^{-n t}\right]\right)^2
}\right)\\
\times \left(2+\frac{1}{\left(1+b e^{n t} \left(B+e^{n t}\right)^{-1+b} n C[1]+2 \text{F1}\left[1,1,1+b,-B e^{-n t}\right]\right)^2}-\frac{3}{1+b e^{n t}
\left(B+e^{n t}\right)^{-1+b} n C_1+2 \text{F1}\left[1,1,1+b,-B e^{-n t}\right]}\right)
\end{array}
\end{equation}

Fig. \ref{fig:6} depicts the evolutionary behaviour of $f(\phi(t))=\mathcal{F}(t)_{NOHDE}$, where we have reconstructed the k-essence model of DE in the cosmological framework of Nojiri-Odintsov holographic fluid. We have observed that the $\mathcal{F}(t)_{NOHDE}$ stays at a positive level. However, if we have a deeper insight into the surface of the 3D plot, we observe that the surface has a significant variation with $n$. For the smaller values of $n$, the surface is not having visible variation with time. However, as $n$ increases, an apparent decreasing pattern of the surface is visible with $t$.

\section{Conclusions}
Based on string theory and black hole thermodynamics, the holographic principle ties a quantum field theory's maximum distance to its infrared cutoff, which is proportional to vacuum energy. The current study investigates a reconstruction scheme for the k-essence form of dark energy using the most generalized version of holographic dark energy introduced by S. Nojiri and S. D. Odintsov (2006) \cite{Odi6} and (S. Nojiri and S. D. Odintsov, 2017, European Physical Journal C, 77, pp.1-8). In this first phase of the work, we start with a reconstruction strategy for the k-essence model using Tsallis holographic Ricci dark energy and then move on to a highly generalized version of holographic dark energy. Initially, we assume that the scale factor is emergent \cite{emergent}. An emergent universe model is fascinating to design, because it has the potential to address a few basic shortcomings in the original Big Bang model. According to \cite{emergent}, the Einstein static universe is an asymptote for time travel. The usual expansion phase begins after this stage, avoiding the big bang singularity. The scaling factor can be expressed as $a(t)=A (e^{nt}+B)^b$ (emergent). In Fig. \ref{fig:1}, we have displayed $X$ as obtained in Eq. (\ref{case1}), which is the holographically reconstructed $X$, with $L=H^{-1}$. This figure displays an expanding trend with redshift $z$. This suggests that, as the cosmos evolves, $X$ increases.  In contrast, Fig. \ref{fig:2}, where we have plotted $X$ versus $z$ based on Eq. (\ref{case2}), the $X$ shows a fading pattern with the evolution of the universe. Using k-essence holography, we considered $\rho_{DE}=\rho(\phi, X)=f(\phi) (-X+3X^2)$. Therefore, $f(\phi)=\frac{\rho_{DE}}{(-X+3X^2)}$. We reconstructed $f(\phi)$ as a function of $t$, represented by $\mathcal{F} (t)$. In Figs. \ref{fig:3} and \ref{fig:4}, we plotted $\mathcal{F} (t)_{case 1}$ and $\mathcal{F} (t)_{case 2}$ with $t$, and in the case of $\mathcal{F} (t)_{case 1}$, we noticed that $\mathcal{F} (t)_{case 1} \rightarrow 0$ as $t \rightarrow 0$. We regarded this as evidence that a sufficient criteria for a realistic reconstruction model had been met. In Fig. \ref{fig:4}, we see that the reconstructed $\mathcal{F} (t)$ initially had a rising pattern with $t$, but at a certain point, it showed a decaying pattern. However, $\mathcal{F} (t)_{case 2} \rightarrow 0$ as $t \rightarrow 0$ was satisfied in this case also. Thus, one of the requirements for a realistic model was met here. In the following phase of the study, we are inspired by \cite{Odi6} that demonstrated that the coincidence problem could be overcome in generalized holographic dark energy, and presented a similar scenario in \cite{Odi7}, where the interaction between matter and holographic energy allows for a constant ratio of matter-energy density to holographic energy.

According to \cite{Odi4,Odi6}, particle and future event horizons and their derivatives influence the holographic cut-off (LIR) \cite{Odi4}  $ L_{IR} = L_{IR}\left(L_p, \dot{L}_p, \ddot{L}_p, ···, L_f, \dot{L}_f, ···,a\right)$, and we considered the generalized model of holographic dark energy with the Nojiri-Odintsov cut-off. Figure \ref{fig:5} plots the EoS parameter for the NO-HDE derived in (\ref{EoSNOHDE}) over a range of $n$ values. The EoS parameter is found to be close to $-1$ for $n$ values near to $0$. However, for bigger $n$ values, the EoS parameter behaves as a quintessence. In Figure \ref{fig:6}, we presented the evolutionary behavior of $f(\phi(t))=\mathcal{F}(t)_{NOHDE}$, where we reconstructed the k-essence model of DE in the cosmic framework of Nojiri-Odintsov holographic fluid. We discovered that $\mathcal{F}(t)_{NOHDE}$ remains positive. However, if we go deeper into the surface of the 3D map, we can see that it varies significantly with $n$. For smaller values of $n$, the surface does not show apparent variation over time. However, as $n$ grows, the surface's apparent decreasing pattern becomes discernible with $t$. Reference \cite{revision2} used Chevallier-Polarski-Linder parametrization to demonstrate that present and future observations favor the CDM scenario over the WDM for dark energy EoS. They limited the current value of the EoS parameter by $-1.06_{-0.13}^{+0.11}$ using observational data sets from SNLS3, BAO, and Planck + WMAP9 + WiggleZ. In a more recent research \cite{revision3}, the present value of DE EoS is confined by observations to be close to $-1$ $(-1.019_{-0.080}^{+0.075})$ according to the $95\%$ limitations imposed by Planck data combined with other astrophysical measurements \cite{revision4}. In view of that, we have compared the EoS parameter obtained in Eq. (\ref{thdew}) with the results of \cite{revision2} and \cite{revision3} in Table \ref{table:1}. It is understandable from the table that for different combinations of values of the model parameters the current values of EoS parameter is in consistency with observational data.

The reference \cite{revision5} suggested a geometrical covariant method for producing holographic hypersurfaces. The covariant-preferred screens for phantom and non-phantom regions are compared to those obtained from the holographic dark energy model, with the infrared cut-off set to the future event horizon. In the context of our study, the outputs of \cite{revision5} appear to be significant. Along with the consistency of the results in the relevant areas, we propose to expand the existing technique to geometrical covariant. Another significant work in this context is by the reference \cite{revision6}, who found a link between the current updated HDE model and the k-essence scalar field model. The model's k-essence potential and scalar field are reconstructed to describe the universe's accelerated expansion phase. Our study, which focused on more generalized variants of HDE, produced results that were consistent with the reference \cite{revision6}. While concluding, let us make some final remarks. NO-HDE is the most generalized version of HDE, and the current work is essentially inspired by \cite{kessence1}. Our study deviates from \cite{kessence1} in the sense that in this study, along with THDE we have considered \cite{Odi4}  $ L_{IR} = L_{IR}\left(L_p, \dot{L}_p, \ddot{L}_p, ···, L_f, \dot{L}_f, ···,a\right)$, which is a highly generalized version of IR cut-off. As future study, we intend to investigate realization of inflation with this kind of scalar field reconstructed through highly generalized IR cut-offs. Holographic Nojiri-Odintsov HDE and Tsallis HDE, which constitute highly specific cases of Nojiri-Odintsov HDE, are studied in this paper. Using the k-essence theory, we reconstructed the associated dark energy. We suggest applying the general approach used here to further specific HDE examples, such as Renyi, Kaniadakis, and other HDEs, which are again particular representatives of Nojiri-Odintsov HDE. The research may also be expanded to include the holographic inflation proposed by Nojiri-Odintsov-Saridakis \cite{revision1}. Additionally, we hope to use fractional-order artificial neural networks in future research to limit the model parameters using observational data. Another possible extension of the current study is to investigate inflation realization with logamediate and intermediate scale factors. 

\section{Acknowledgement}
The authors sincerely acknowledge the constructive suggestions from the anonymous reviewers. This research was funded by the Science Committee of the Ministry of Science and Higher Education of the Republic of Kazakhstan (Grant No. AP14972654).

\end{document}